\documentclass[twocolumn,aps,prb,showpacs]{revtex4-1}
\usepackage{graphicx,psfrag}
\usepackage{graphics,psfrag}
\usepackage{amssymb,amsfonts,amsmath}
\usepackage{bbm}
\usepackage{dcolumn}
\usepackage{color}
\usepackage{subfig}
\usepackage{rotating}
\usepackage{verbatim}
\usepackage[toc,page]{appendix}

\def\xa{{x_{\alpha}}}
\def\xb{{x_{\beta}}}
\def\xc{{x_{\gamma}}}
\def\xd{{x_{\gamma}}}

\def\yac{{y_{\alpha\gamma}}}

\def\zabcd{{z_{\alpha\beta\gamma\delta}}}
\def\abcd{{\alpha\beta\gamma\delta}}
\def\kB{{k_{\rm B}}}

\def\SPoint{{S_{\rm Point}}}
\def\SNN{{S_{\rm NN}}}
\def\SNNN{{S_{\rm NNN}}}
\def\STri{{S_{\rm Tri}}}
\def\STetra{{S_{\rm Tetra}}}
\def\bP{{\bf P}}
\def\br{{\bf r}}
\def\rmd{{\rm d}}
\def\blp{\bigl(} \def\brp{\bigr)}
\def\avg#1{{\langle #1\rangle }}

\begin{document}

\title{Interaction models and configurational entropies of binary MoTa and the MoNbTaW high entropy alloy}
\author{Andrew D. Kim and Michael Widom}
\affiliation{Department of Physics, Carnegie Mellon University, Pittsburgh PA 15213}

\date{\today}

\begin{abstract}
We introduce a simplified method to model the interatomic interactions of high entropy alloys based on a lookup table of cluster energies. These interactions are employed in replica exchange Monte Carlo simulations with histogram analysis to obtain thermodynamic properties across a broad temperature range. Kikuchi's Cluster Variation Method entropy formalism and high temperature series expansions are applied to directly calculate entropy from statistics on short- and long-range chemical order, and we discuss the convergence of the entropy as clusters of differing size are included.  
\end{abstract}

\pacs{}

\maketitle

\section{Introduction}

High entropy alloys (HEAs) are multicomponent solid solutions that randomly distribute chemical species among the sites of a crystal lattice~\cite{Yeh04_1,Cantor04}. Although the distribution is nominally random, preferences in chemical bonding correlate the chemical identities of nearby atoms, creating short-range chemical order that reduces the entropy below the ideal value of $\kB\ln{(N)}$ for $N$ species. In the case of strong bonding preferences, or in equilibrium at low temperatures, long-range chemical order and even phase separation may arise~\cite{deFontaine79,Huhn13,Widom2015}.

Computer simulation provides powerful techniques to predict and quantify chemical order for a given model of interatomic interaction. Models range from accurate but expensive first principles methods, through reasonably accurate though complicated machine learning and cluster expansion approaches, to simplified empirical formulas such as embedded atom or pair potentials. Here we introduce a simple but accurate approach based on a lookup table of precalculated first principles energies. We then apply replica exchange Monte Carlo simulation~\cite{Swendsen86} to reach equilibrium over a broad range of temperatures.

The resulting data set allows us to calculate thermodynamic quantities, including the entropy, through use of the multiple histogram method~\cite{Ferrenberg88,Ferrenberg89}. We compare our results with formulas adapted from the Cluster Variation Method~\cite{Kikuchi1951,Ackermann89} (CVM) that express the entropy in terms of simulated cluster probabilities, and show the convergence of the CVM entropy with respect to the included clusters. The pattern of convergence is interpreted through the use of a high temperature series expansion~\cite{KardarFields} that confirms the sequence of optimal clusters~\cite{vul_fontaine_1992}.

Our approach is illustrated through application to the widely studied body centered cubic (BCC) refractory MoNbTaW high entropy alloy~\cite{Senkov10}. This compound is believed to exhibit strong short-range chemical order in equilibrium (although this is difficult to achieve experimentally) due to strong binding of Mo and Ta at nearest neighbors. Our model exhibits the expected Strukturbericht A2 to B2 (Pearson type cI2 to cP2) ordering transition at intermediate temperatures, and phase separation at low temperatures~\cite{Huhn13,Wrobel,Shapeev,Ikeda}. Finite size variation of the specific heat and susceptibility peaks indicate that the transition is of the 3D Ising type. We compare the behavior of the 4-component HEA with the binary solid solution MoTa, which provides a simpler picture of similar behavior.

\section{Methods}

This section describes our interaction model, the replica exchange simulation method, histogram analysis of simulation data, the cluster variation method entropy formulas, and the high temperature series expansion. Several of our codes are available at~\cite{euler-MC}.

\subsection{Interaction Model}

\begin{figure}[!h]
  \includegraphics[width=.49\textwidth]{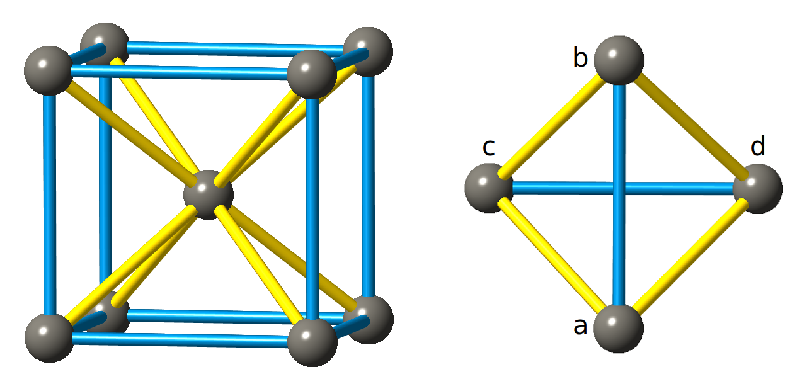}
  \caption{\label{fig:BCC-tetra} BCC unit cell (left) and BCC tetrahedron (right). Nearest neighbor bonds are in yellow and next-nearest neighbor bonds are in blue. Tetrahedron sites $\alpha$ and $\beta$ are ``even'' (cube vertex) while $\gamma$ and $\delta$ are ``odd'' (body center) sites.}
\end{figure}

Each of the four chemical species (Mo, Nb, Ta and W) individually take the BCC crystal structure. In combination they occupy sites of the BCC lattice to form a disordered solid solution at high temperatures, but they order and eventually phase separate at low temperatures, all the while maintaining the underlying BCC sites. We shall be interested in clusters containing nearest and next-nearest neighbor pairs, including triangles and the BCC tetrahedron (see Fig.~\ref{fig:BCC-tetra}). This four-point cluster may be considered as the primitive cell of the quaternary Heusler crystal type~\cite{Hoffman2021} (Strukturbericht L2$_1$, Pearson cF16), and hence repeated periodically to fill space.

We enumerated the complete set of $4^4=256$ arrangements of the four species on the four tetrahedron sites, of which $55$ are symmetry-inequivalent, and calculated the corresponding Heusler structure energies using density functional theory (DFT). Specifically, we employ {\tt VASP}~\cite{Kresse96} with projector augmented wave potentials~\cite{Kresse99} in the PBE generalized gradient approximation~\cite{Perdew96}. The cubic symmetries of the structures prevented atomic relaxation, and we held the lattice parameters fixed at 3.2305~\AA.

The resulting energies provide a lookup table that can be used to quickly evaluate the energy of any arrangement of the chemical species on BCC lattice sites. To evaluate the total energy, we decompose the structure into its constituent tetrahedra. Let $\alpha$, $\beta$, $\gamma$ and $\delta$ designate the chemical species at tetrahedron vertices $a$, $b$, $c$ and $d$, respectively. This tetrahedron contributes energy $E(\alpha,\beta,\gamma,\delta)/24$ per atom to the total energy, where $E(\alpha,\beta,\gamma,\delta)$ is the energy per primitive cell of the 4-atom Heusler crystal with species $\alpha$, $\beta$, $\gamma$ and $\delta$. An additional factor of 6 arises because the BCC structure has 6 tetrahedra per atom. Our energies are calculated relative to the average energies of elemental Mo, Nb, Ta and W, so they represent energy of formation. We may think of this energy as a cluster expansion~\cite{ATAT2002} containing just a single cluster.

\begin{figure}[!h]
  \includegraphics[width=.49\textwidth]{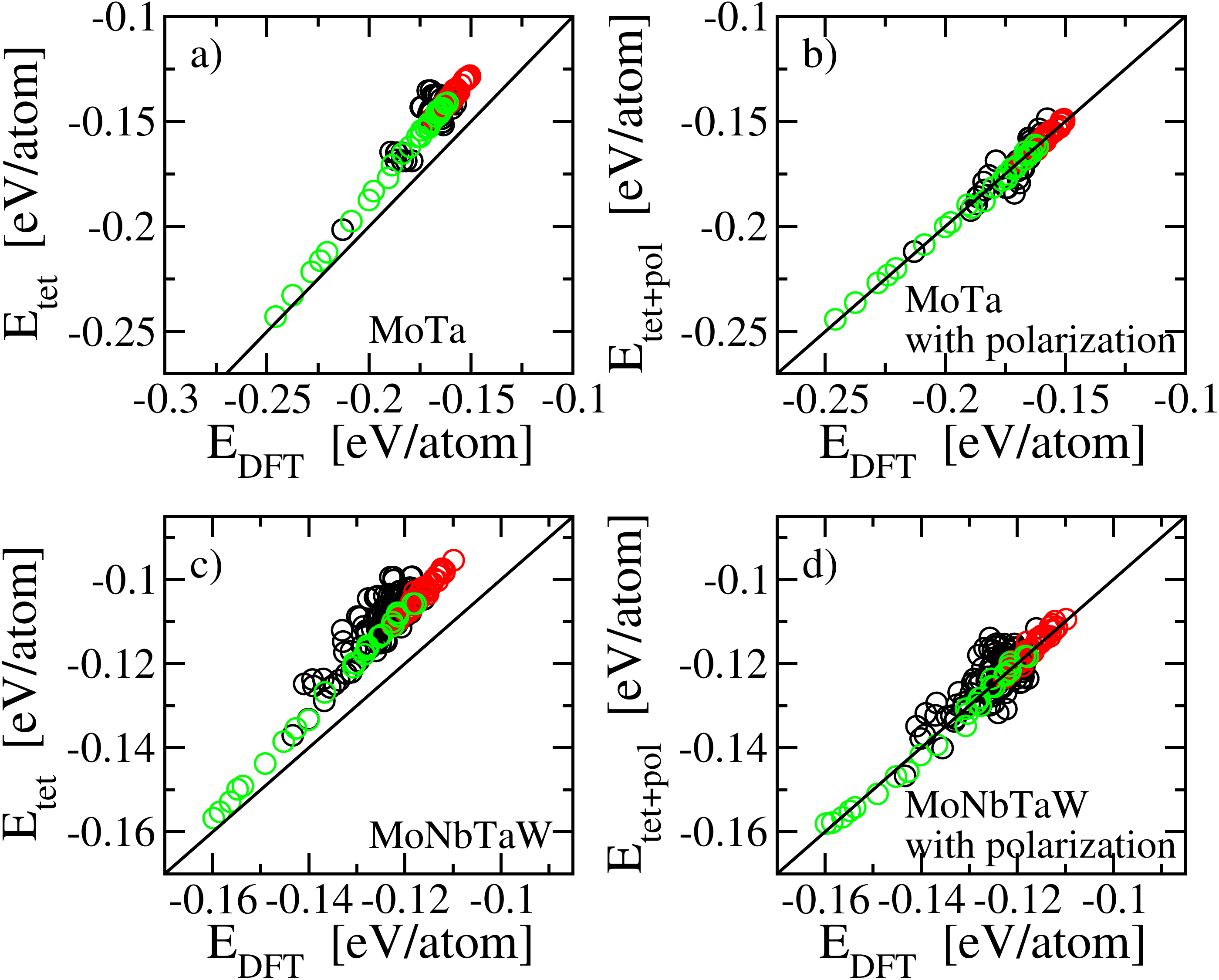}
  \caption{\label{fig:ParityPlots} Parity plot for model energy {\em vs.} full DFT energy for MoTa ((a) and (b)) and MoNbTaW ((c) and (d)). Plots (a) and (c) are based on tetrahedron energies only, while (b) and (d) include polarization corrections. All structures are equiatomic; black are random 16-atom, red are random 128-atom, and green are MC-generated 128-atom structures.}
\end{figure}

To test the accuracy of this model we created parity plots of model energy {\em vs.} full DFT energy as shown in Fig.~\ref{fig:ParityPlots}(a) for equiatomic binary MoTa and (c) for equiatomic quaternary MoNbTaW.  The model energies lie close to parity with a mean absolute error of $21$ meV/atom for MoTa and $14$ meV/atom for MoNbTaW. Crucially, the lowest energy structures lie very close to parity, so we accurately capture the low temperature properties. 

We observe a systematic skewing of model energy above the parity line, suggesting that the structures for our tabulated values are systematically higher than our randomly generated test sets. In fact, since our lookup table is based on primitive cells of the cubic Heusler structure, every atom is in an environment of perfect cubic symmetry, while in a random structure most atoms are in environments of low symmetry. Anisotropic charge transfer can create local electric fields that will polarize the atoms. Hence, we proposed a correction to the energy of the form
\begin{equation}
  \label{eq:polar-dE}
  \Delta E = -\frac{|\bP|^2}{2\chi_p}
\end{equation}
where $\chi_p$ is an adjustable parameter related to dielectric susceptibility and the ``polarization''
\begin{equation}
  \label{eq:polar-P}
  \bP = \sum_\br \br \chi_e(\br)
\end{equation}
is a measure of the anisotropy of Pauling's electronegativity $\chi_e$. This one-parameter correction results in improved agreement with a mean absolute error of $1$ relative to the full DFT energies for MoTa and $3$ for MoNbTaW.

\subsection{Replica exchange simulation}

Replica exchange simulations~\cite{Swendsen86}, also known as parallel tempering, aim to accelerate the sampling of configuration space by sharing multiple configurations (replicas) among multiple temperatures, in a manner that preserves the properly weighted ensemble at each temperature. When configurations are swapped between temperatures, the diversity of the equilibrium ensemble at each temperature is enriched by the addition of a new independent configuration. From the perspective of a single configuration, getting swapped to a higher temperature may facilitate its evolution by raising the likelihood of escape from a local energy minimum.

\begin{figure}[!h]
  \includegraphics[width=0.49\textwidth]{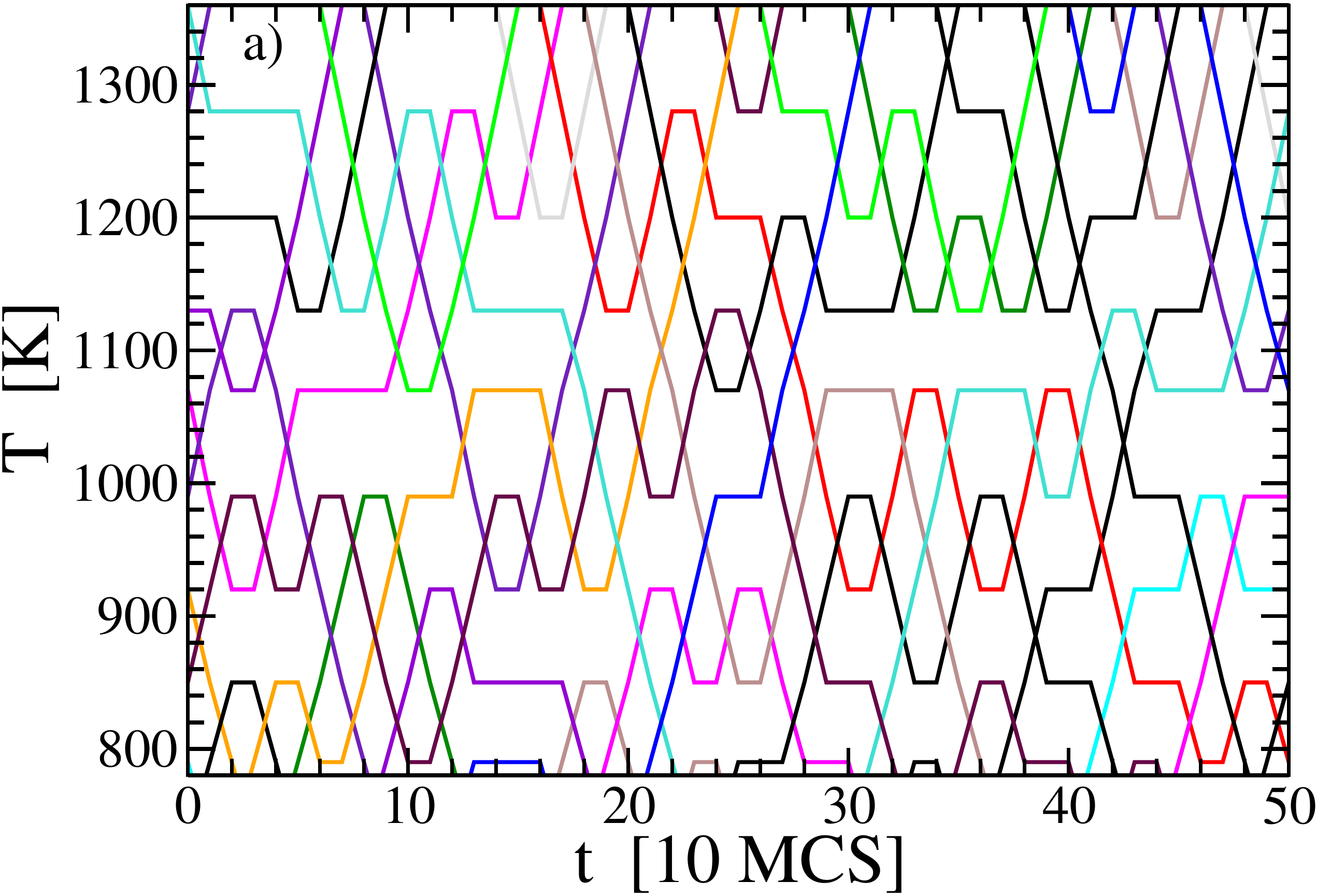}
  \includegraphics[width=0.49\textwidth]{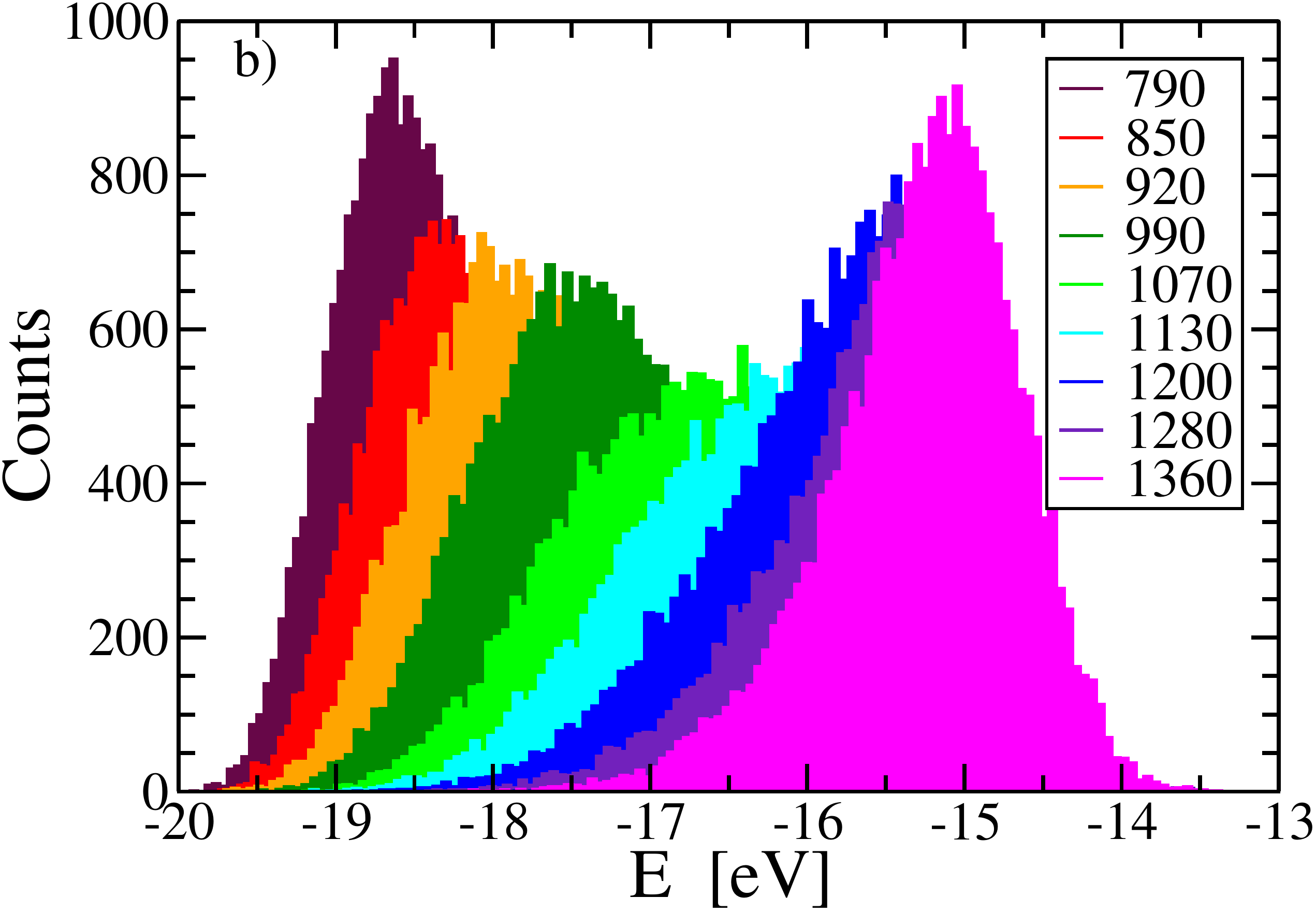}
  \caption{\label{fig:Tvst} Illustration of quaternary MoNbTaW replica exchange simulation. (a) T(t) graph illustrating replica exchange; (b) energy histograms at selected temperatures.}
\end{figure}

The probability for a configuration $C_i$ of energy $E_i$ to occur in equilibrium at temperature $T$ (inverse temperature $\beta=1/\kB T$) is $P_i=\exp{(-\beta E)}/Z(T_i)$, with $Z(T)$ the partition function for temperature $T$. The joint probability for a pair of configurations $C_i$ of energy $E_i$ and $C_j$ of energy $E_j$ at respective temperatures $T_i$ and $T_j$ is
\begin{equation}
  \label{eq:Pjoint}
  P(E_i,E_j|T_i,T_j) =   e^{-(\beta_i E_i + \beta_j E_j)}/Z(T_i)Z(T_j)
\end{equation}
The joint probability for $C_i$ to occur at temperature $T_j$, and $C_j$ to occur at temperature $T_i$ is given by the same formula with energies $E_i$ and $E_j$ interchanged. The ratio of probabilities
\begin{equation}
  \label{eq:Pratio}
  P(C_i,C_j|T_i,T_j)/P(C_j,C_i|T_i,T_j)
  = e^{\Delta\beta \Delta E}
\end{equation}
with $\Delta\beta=\beta_i-\beta_j$ and $\Delta E = E_i-E_j$. Hence, the equilibrium ensemble probabilities at temperatures $T_i$ and $T_j$ are preserved if the configurations are swapped with the probability given by the ratio in Eq.~(\ref{eq:Pratio}). Note that, conveniently, the partition functions (which are usually unknown in computer simulations) cancel in the ratio.

In addition to replica swaps we must evolve the configurations at each temperature. We apply a Monte Carlo process to attempt discrete swaps of chemical species. Equilibrium at temperature $T$ is maintained if the attempts are accepted with probability $\exp{(-\Delta E/\kB T)}$ with $\Delta E$ the energy change created by the swap.

Figure~\ref{fig:Tvst}a illustrates a portion of the time evolution of thermostat temperature. Each configuration is represented by a different color. Temperatures are spaced so that potential energy distributions of adjacent temperatures overlap sufficiently (see Fig.~\ref{fig:Tvst}b) to achieve temperature swap probabilities of 20\% or greater. Data collection began following lengthy pre-annealing during which energy and order parameters distributions relaxed to steady states. By calculating averages and fluctuations of energy at each temperature, we obtain the energy and specific heat at the simulated temperatures. We also gather statistics on the frequencies of occurrence, $\zabcd$, for tetrahedra with chemical species $\abcd$.

\subsection{Histogram analysis}

The probability that any configuration has energy $E$ in equilibrium at temperature $T$ is $P(E)=\Omega(E)\exp{(-\beta E)}/Z(T)$, where $\Omega(E)$ is the configurational density of states. If we knew $\Omega(E)$, we could evaluate the partition function
\begin{equation}
  \label{eq:Z}
  Z(T)=\int\rmd E~ \Omega(E)~ e^{-E/\kB T},
\end{equation}
and then obtain the free energy as $F=\kB T\ln{Z}$. Given $F(T)$, we may obtain the internal energy as $U=-\partial (\beta F)/\partial\beta$ and the specific heat as $c=\partial U/\partial T=-T\partial^2 U/\partial T^2$.

During a simulation at temperature $T$ the frequency with which energy $E$ occurs is proportional to the probability $P(E)$. Hence the density of states $\Omega(E)$ is proportional to a histogram of energies, $H_T(E)$, up to the factor $\exp{(-\beta E)}/Z(T)$. Note that the normalizing partition function $Z(T)$ is unknown, but it is {\em independent} of the energy $E$. We may express~\cite{Ferrenberg88}
\begin{equation}
  \label{eq:Hist}
  \Omega_T(E)=H_T(E)~ e^{\beta E},
\end{equation}
where the subscript $T$ reminds us that this is an approximation obtained from a simulation at temperature $T$, and that it differs from the true $\Omega$ by an unknown constant factor. Substituting $\Omega_T$ into Eq.~\ref{eq:Z} and taking the logarithm, we can evaluate the free energy $F(T')$ as a continuously varying function of temperature $T'$. The unknown constant factor in $\Omega_T$ results in an unknown additive constant in $\beta' F(T')$, however this does not impact the internal energy $U$ and specific heat $c$. The derivatives to obtain $U$ and $c$ may be taken analytically, and their values also become continuously varying functions of $T'$.

The free energy $F(T')$ is most accurate for $T'$ in the vicinity of the simulated temperature $T$ because the histogram is best resolved over the range of highly probable energies. Luckily, the method generalizes to include multiple histograms accumulated at different temperatures $T$, with~\cite{Ferrenberg89}
\begin{equation}
  \label{eq:MultiHist}
  \Omega(E)=\frac{\sum_T H_T(E)}{\sum_T e^{(F(T)-E)/\kB T}},
\end{equation}
where the free energies $F(T)$ must be obtained self-consistently with $\Omega(E)$ through Eqs.~(\ref{eq:Z}) and~(\ref{eq:MultiHist}) for the set of temperatures simulated. In this manner the relative free energy can be extended across the entire range of simulated temperatures. 

\subsection{CVM}

We exploit the formalism of Kikuchi's cluster variation method to define a hierarchy of approximate entropy models based on the sequence of single-, two-, three- and ultimately four-point cluster frequencies. Starting with the four-point frequencies $\zabcd$ already introduced, we define three-point frequencies $u_{\alpha\gamma\delta}$, $u_{\beta\gamma\delta}$, $u_{\alpha\beta\gamma}$, and $u_{\alpha\beta\delta}$ by summing $\zabcd$ over one of its four indices. Similarly we introduce pairs for nearest neighbors, $y_{\alpha\gamma}$, $y_{\alpha\delta}$, $y_{\beta\gamma}$, and $y_{\beta\delta}$; next nearest neighbors $v_{\alpha\beta}$ and $v_{\gamma\delta}$; and points $x_\alpha$, $x_\beta$, $x_\gamma$, and $x_\delta$.

The entropy associated with a given cluster $\Gamma$ is $\Sigma(\Gamma)=-\sum\Gamma\ln{\Gamma}$, where the sum is over the cluster variables. For example,
\begin{equation}
  \label{S_xa}
  \Sigma(\xa) = -\sum_\alpha \xa \ln{\xa}.
\end{equation}
We also introduce shorthand notation $\Sigma(X)=(\Sigma(\xa)+\Sigma(\xb)+\Sigma(\xc)+\Sigma(\xd))/4$, $\Sigma(Y)=(\Sigma(y_{\alpha\gamma})+\Sigma(y_{\alpha\delta})+\Sigma(x_{\beta\gamma})+\Sigma(x_{\beta\delta}))/4$, $\Sigma(V)=(\Sigma(v_{\alpha\beta})+\Sigma(v_{\gamma\delta}))/2$, $\Sigma(U)=(\Sigma(u_{\alpha\gamma\delta})+\Sigma(u_{\alpha\beta\gamma})+\Sigma(u_{\beta\gamma\delta})+\Sigma(u_{\beta\gamma\delta}))/4$, and $\Sigma(Z)=\Sigma(\zabcd)$. As discussed in~\cite{Ackermann89,Yedidia2005,Pelizzola2005,Widom16,Hoffman2021}, we build up higher approximations to the entropy through inclusion of entropy-reducing information contained in successively larger clusters, while correcting for the overcounting of subclusters. Specifically, we obtain
\begin{align}
  \label{eq:S_clusters}
  \SPoint &= \Sigma(X) \\ \nonumber
  \SNN &= -7~\Sigma(X)+4~\Sigma(Y)\\ \nonumber
  \SNNN &= -13~\Sigma(X)+4~\Sigma(Y)+3~\Sigma(V)\\ \nonumber
  \STri &= 23 ~\Sigma(X) - 20 ~\Sigma(V) - 9 ~\Sigma(Y) + 12 ~\Sigma(U)\\ \nonumber
  \STetra &= -\Sigma(X) + 4 ~\Sigma(Y) + 3 ~\Sigma(V) - 12 ~\Sigma(U) + 6 ~\Sigma(Z). \nonumber
\end{align}
Coefficients of the highest order cluster equal the numbers of such clusters per site, while the lower order coefficients reflect the systematic exclusion of subclusters. For example, $\SPoint$ is the Bragg-Williams ideal mixing entropy~\cite{Bragg34}, while $\SNN$ reduces $\SPoint$ by the mutual information contained in the nearest-neighbor cluster frequencies~\cite{Hoffman2021}.

\subsection{High T expansion}
\label{sec:high-T}

As a test of the accuracy of the CVM entropy formulas, we apply it to the BCC nearest neighbor Ising model. The Hamiltonian $H=J\sum_{<ij>}\sigma_i\sigma_j$ is used to evaluate multi-point correlation functions in the high temperature limit~\cite{KardarFields}. Here $J$ is the nearest neighbor coupling constant, and $\sigma_i=\pm 1$ is the spin at site $i=a-d$ (see Fig.~\ref{fig:BCC-tetra}b for site labels). To evaluate Eq.~(\ref{eq:S_clusters}) we need correlation functions up to four-point $z_{\sigma_a\sigma_b\sigma_c\sigma_d}$ expanded up to 4$^{th}$ order in $t\equiv\tanh{(J/k_BT)}$,
\begin{widetext}
\begin{align}
  \label{eq:highT-P4}
  \nonumber
Z~z_{\sigma_a\sigma_b\sigma_c\sigma_d} = \cosh^{4N}{(J/k_BT)}2^{N-4} 
\blp 1 + t (\sigma_a\sigma_c+\sigma_b\sigma_c+
\sigma_a\sigma_d+\sigma_b\sigma_d)+ t^2 (4(\sigma_a\sigma_c+\sigma_b\sigma_d)+ 2\sigma_a\sigma_b\sigma_c\sigma_d) \\
+ t^3  12 (\sigma_a\sigma_c+\sigma_b\sigma_c+\sigma_a\sigma_d+\sigma_b\sigma_d) 
+ t^4 (12 N + 56(\sigma_a\sigma_b+\sigma_c\sigma_d)+12\sigma_a\sigma_b\sigma_c\sigma_d)+\cdots \brp. 
\end{align}
\end{widetext}

Fewer-point correlation functions are obtained by summing over spins, including the partition function itself
\begin{align}
Z &= \sum_{\sigma_a\sigma_b\sigma_c\sigma_d}
     Z~z_{\sigma_a\sigma_b\sigma_c\sigma_d} \\
  &= \cosh^{4N}{(J/k_BT)}2^{N-4} \blp 1+12 N t^4 +\cdots \brp .
\end{align}
Setting the free energy $F=-k_BT \ln{Z}$ and expanding in powers of $1/T$ we obtain entropy per site
\begin{equation}
S/k_B = \ln{2}-2\left(\frac{J}{k_B T}\right)^2-35\left(\frac{J}{k_B T}\right)^4 +\cdots .
\end{equation}
The correlation functions and the entropy are exact up to 4$^{th}$ order and consistent with prior results~\cite{Montroll1953,Baker1963}.

Inserting the correlation functions obtained from the expansion Eq.~(\ref{eq:highT-P4}) into the CVM formulae we obtain the following expansions for the entropy
\begin{align}
S_{\rm NN} &= \ln{2}-2\left(\frac{J}{k_B T}\right)^2-47\left(\frac{J}{k_B T}\right)^4+\cdots\\
S_{\rm NNN} &= \ln{2}-2\left(\frac{J}{k_B T}\right)^2-79\left(\frac{J}{k_B T}\right)^4+\cdots\\
S_{\rm TRI} &= \ln{2}-2\left(\frac{J}{k_B T}\right)^2-29\left(\frac{J}{k_B T}\right)^4+\cdots\\
S_{\rm TET} &= \ln{2}-2\left(\frac{J}{k_B T}\right)^2-35\left(\frac{J}{k_B T}\right)^4+\cdots .
\end{align}
Notice that we obtain the correct quadratic term already using simply the nearest-neighbor correlation function $\yac$. However the quartic term is too large in magnitude, so that the NN approximation underestimates the entropy as temperature drops. This overcorrection is due to the presence of closed loops of NN bonds causing the same information to be counted multiple times. For example, the correlation propagated from $a$ to $d$ passing through $c$ and $b$ augments the direct correlation of $a$ with $d$ (see Fig.~\ref{fig:BCC-tetra}b). Including the NNN term makes the problem worse, because we are subtracting the mutual information between $a$ and $b$ yet again. Inclusion of the TRI term overcompensates and consequently overestimates the entropy, while, finally, inclusion of the TET term restores the proper quartic coefficient.

\section{Results}

\subsection{MoTa}

\begin{figure}[h!]
  \includegraphics[width=.49\textwidth]{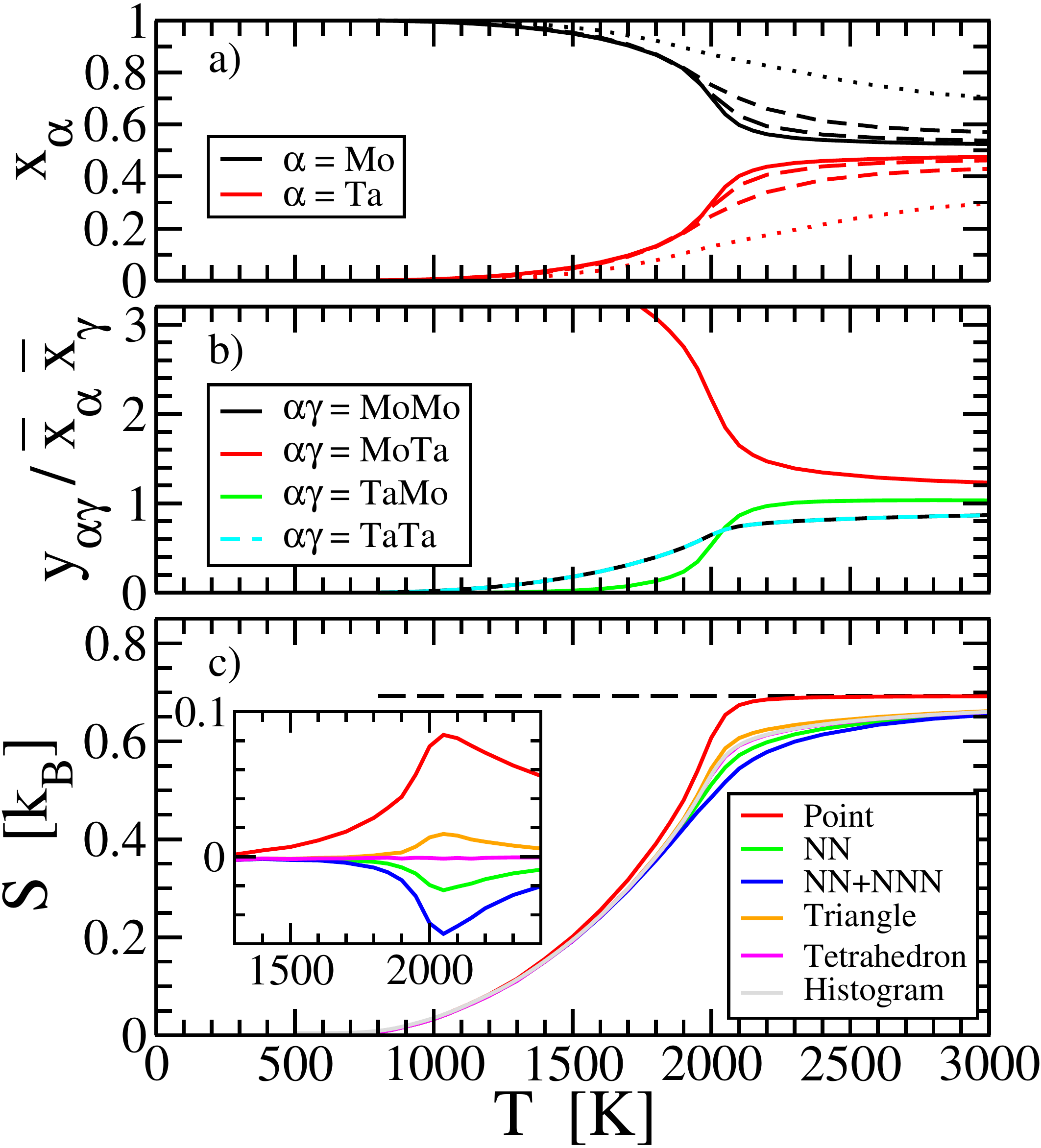}
  \caption{\label{fig:MoTa-entropy} Binary MoTa of size $L=8$ (1024 atoms). (a) Occupation $\xa$ of even sites $a$ and $b$ (dots, short, and long dashes are sizes $2$, $4$, and $6$, respectively); (b) $ac$ site pair frequencies $\yac$ normalized by global mean concentrations $\bar{x}_{\alpha}$ and $\bar{x}_{\gamma}$; (c) Simulated histogram and CVM-predicted entropies (inset: residuals with respect to histogram). Dashed line shows $S/\kB=\ln{2}$}
\end{figure}

Figure~\ref{fig:MoTa-entropy}a plots the temperature-dependent $a$-site occupation $\xa$ over the range from 800K up to the approximate melting temperature of 3000K. We adopt a convention where we shift the simulated structure so that the maximum Mo occupation occurs on the ``even'' sublattice (site classes $a$ and $b$). Hence, a slight bias artificially raises $x_{\rm Mo}$ relative to $x_{\rm Ta}$ for finite system sizes. Evidently a transition to long-range order occurs in the vicinity of $T_c\approx 2020$K, with $\xa$ converging towards the global mean concentration $\bar{x}_\alpha=1/2$ above $T_c$ but diverging away below. Even above $T_c$ the NN pair frequencies $\yac$ deviate from the independent expectation $\xa\xc$, as shown in Fig.~\ref{fig:MoTa-entropy}b, with an enhanced frequency of MoTa pairs. The discrepancy grows rapidly below $T_c$.

The deviation of $\yac$ from independence above $T_c$, and the additional symmetry-breaking of the site occupation below $T_c$, cause the entropy to fall below its ideal mixing value of $\kB\ln{2}$. As shown in Fig.~\ref{fig:MoTa-entropy}c, the entropy loss accelerates below $T_c$, and the net entropy tends towards zero at low temperatures. We show CVM cluster-based estimates of entropy and our histogram values. The histogram method yields only relative entropy, so we adjust it to match the CVM tetrahedron value at 3000K; the fact that it vanishes at low temperature indicates its accuracy across the full temperature range.

The inset shows entropy residuals relative to the histogram method. Note that the point values under-correct the ideal mixing, while the pair value over-corrects and the two-pair strongly over-corrects. As seen in our discussion of the high temperature series (Sect.~\ref{sec:high-T}), this can be attributed to the cumulative effect of correlations extending around closed loops. The deviations are maximal around $T_c$, supporting the role of longer-range correlations. The triangle approximation overcorrects in the opposite direction, while the tetrahedron values lie close to the multiple histogram at all temperatures.

\begin{figure}[!h]
  \includegraphics[width=.49\textwidth]{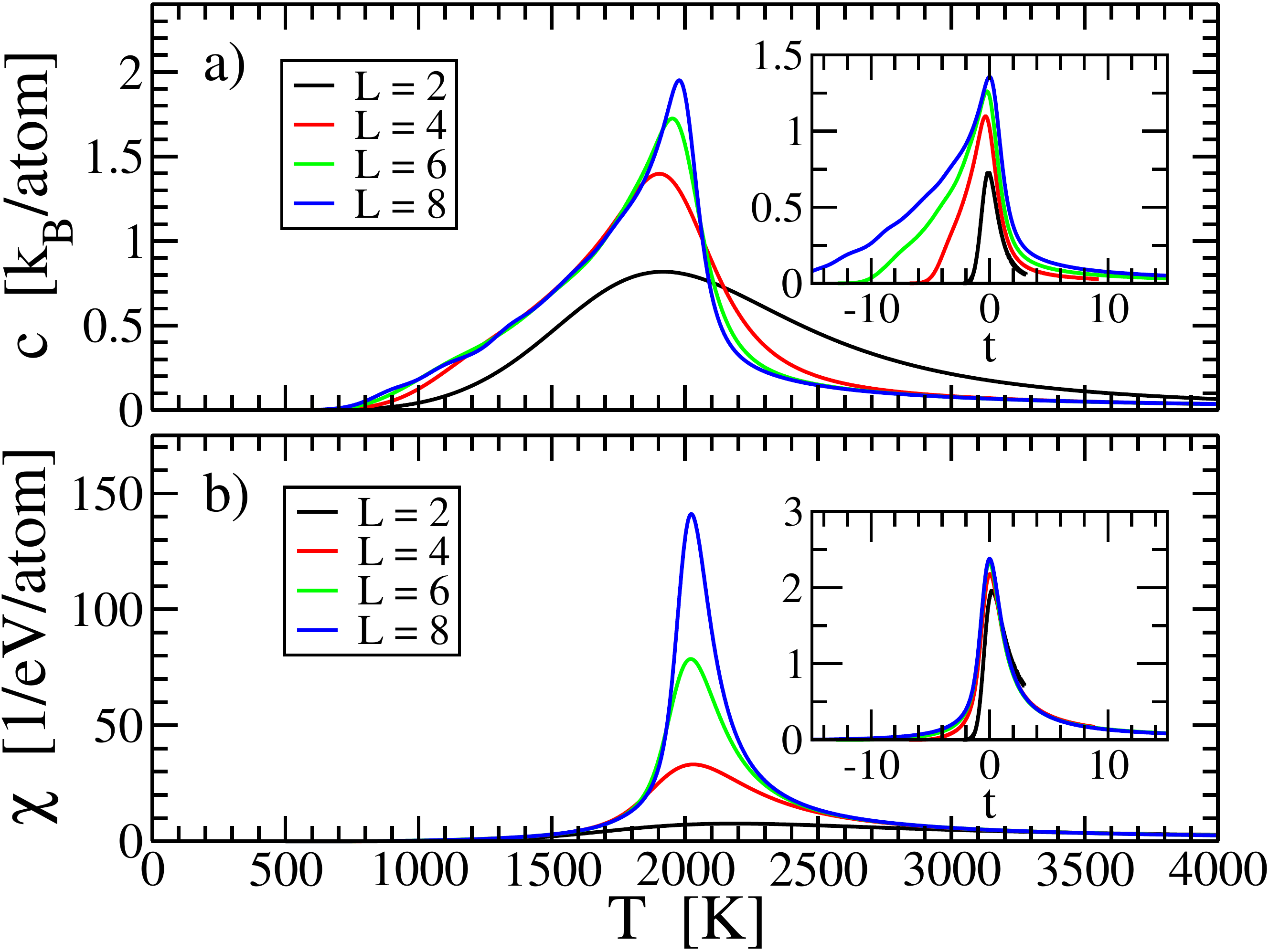}
  \caption{\label{fig:MoTa-sim} Binary MoTa. (a) Specific heat $c(T)$, and (b) Susceptibility $\chi(T)$ {\em vs.} temperature $T$ for system sizes $L=2$ (black), $4$ (red), $6$ (green) and $8$ (blue). Insets are scaling functions as defined in text.}
\end{figure}

In order to confirm the presence of a genuine thermodynamic phase transition, we plot the specific heat and generalized susceptibility in Fig.~\ref{fig:MoTa-sim}. The generalized susceptibility is defined as
\begin{equation}
  \label{eq:chi}
  \chi=\frac{N_s}{\kB T}\left(\avg{M^2}-\avg{M}^2\right)
\end{equation}
where we define $M=x_{\rm Mo}-x_{\rm Ta}$ on the even sites, and $N_s$ is the number of sites.  Divergences in $c$ and $\chi$ reveal thermodynamic singularities. In order to determine the character of the phase transition, we plot the scaled specific heat $c/L^{\alpha/\nu}$ and the scaled susceptibility $\chi/L^{\gamma/\nu}$ as functions of the scaled reduced temperature $t=L^{1/\nu}(T-T_c)/T_c$ in the insets. We set $T_c$ as the peak susceptibility temperature for $L=8$, and we take the established 3D Ising values of the critical exponents $\alpha=0.110$, $\gamma=1.2372$, and $\nu=0.6301$. The convergence towards common scaled functions indicates the transition is in the Ising class, as  expected for a binary alloy. 

The low temperature phase takes the CsCl structure (Strukturbericht type B2, Pearson cP2). The actual ground state according to full first principles calculations~\cite{Huhn13} is Pearson type oC12, which is locally B2 with periodic antiphase faults; the B2 phase lies above oC12 by just 1 meV/atom. A total of 10 distinct ordered phases are predicted at varying compositions, all based on an underlying BCC lattice. Experimentally, only a solid solution is reported, with no ordered phases.

\subsection{MoNbTaW}

\begin{figure}[!h]
    \includegraphics[width=.49\textwidth]{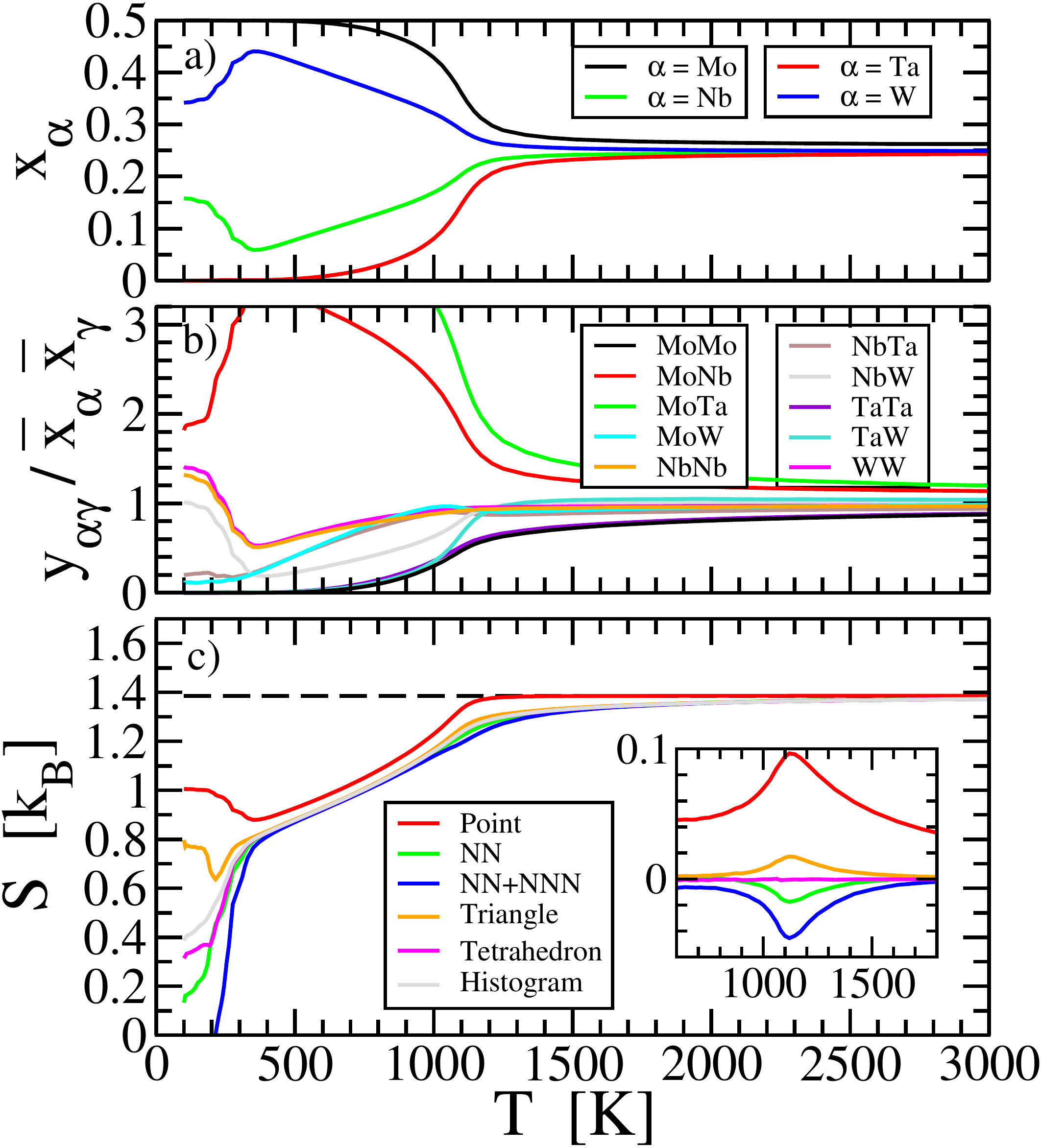}
  \caption{\label{fig:MoNbTaW-entropy} Quaternary MoNbTaW of size $L=8$ (1024 atoms). (a) Occupation $\xa$ of $a$ site; (b) $ac$ site pair frequencies $\yac$ normalized by global mean concentrations $\bar{x}_{\alpha}$ and $\bar{x}_{\gamma}$; (c) Simulated histogram and CVM-predicted entropies (inset: residuals with respect to histogram). Dashed line shows $S/\kB=\ln{4}$.}
\end{figure}

\begin{figure}[!h]
  \includegraphics[width=.49\textwidth]{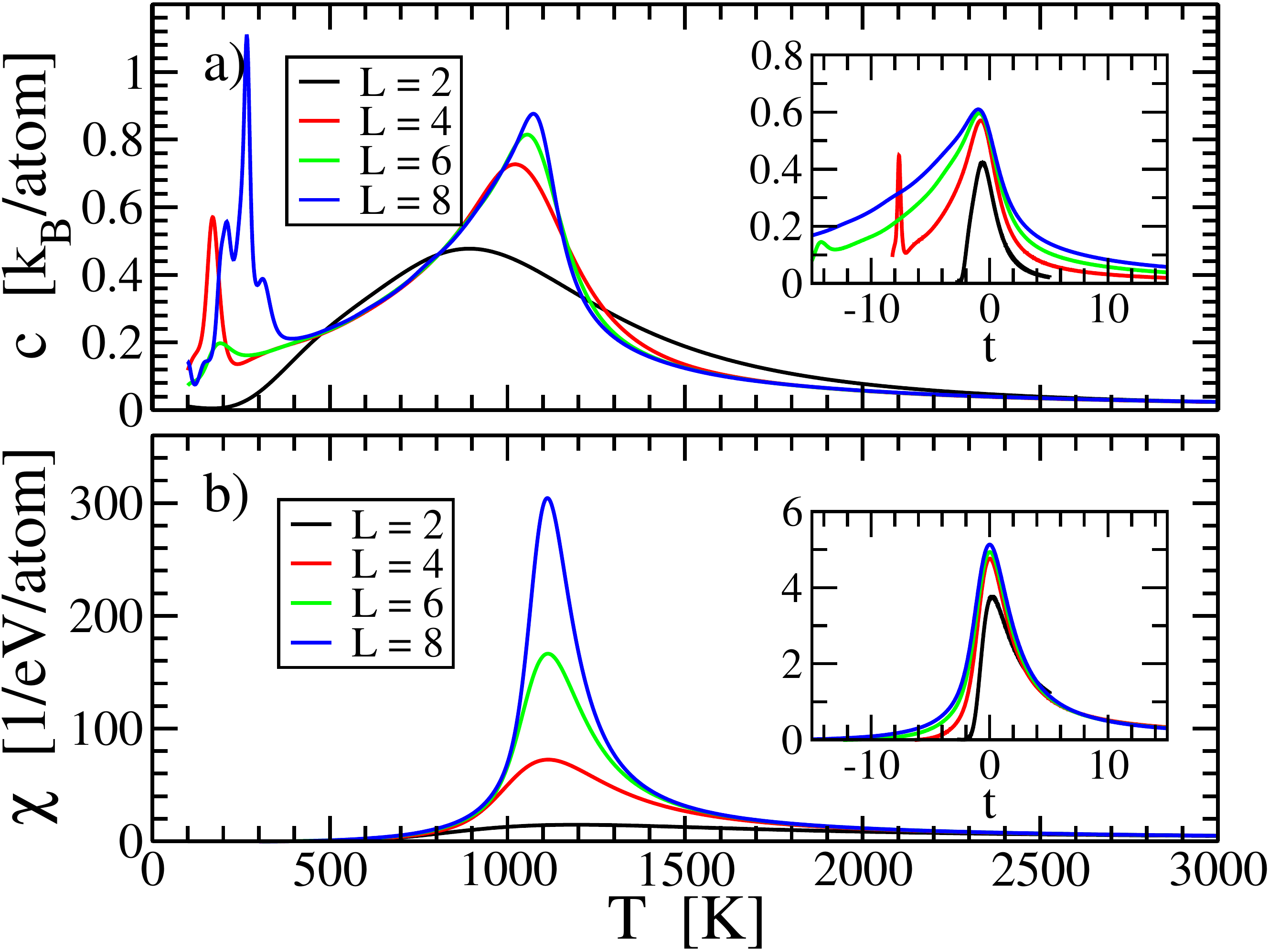}
  \caption{\label{fig:MoNbTaW-sim} Quaternary MoNbTaW. (a) specific heat $c(T)$, and (b) Susceptibility $\chi(T)$ {\em vs.} temperature $T$ for system sizes $L=2$ (black), $4$ (red), $6$ (green) and $8$ (blue). Insets are scaling functions as defined in text.}
\end{figure}

The MoNbTaW quaternary behaves similarly to the MoTa binary. As seen in Fig.~\ref{fig:MoNbTaW-entropy}, the dominant ordering occurs between Mo and Ta, on the even and odd sites, respectively. The Nb occupation and correlations generally follow Ta, and W generally follows Mo, as previously seen~\cite{Huhn13}. Likewise, the accuracy of the CVM entropy estimates closely resemble the binary case. The specific heat and susceptibility also show an Ising-like transition (Fig.~\ref{fig:MoNbTaW-sim}), though at a lower critical temperature of $T_c\approx 1110$K compared with the binary case. The lower critical temperature can be attributed to dilution of the strongly interacting MoTa pairs by the more weakly interacting Nb and W. Interestingly, Mo and Ta show relatively strong variation in the electronegativity and atomic volume as compared with Nb and W - electronegativity: Ta (1.5) $\lesssim$ Nb (1.6) $\ll$ Mo (2.16) $\lesssim$ W (2.36) in Pauling units; atomic volume: Ta (18.00) $\gtrsim$ Nb (17.97) $\gg$ W (15.85) $\gtrsim$ Mo (15.55) in \AA$^3$/atom.

\begin{figure}[!h]
  \includegraphics[width=.49\textwidth]{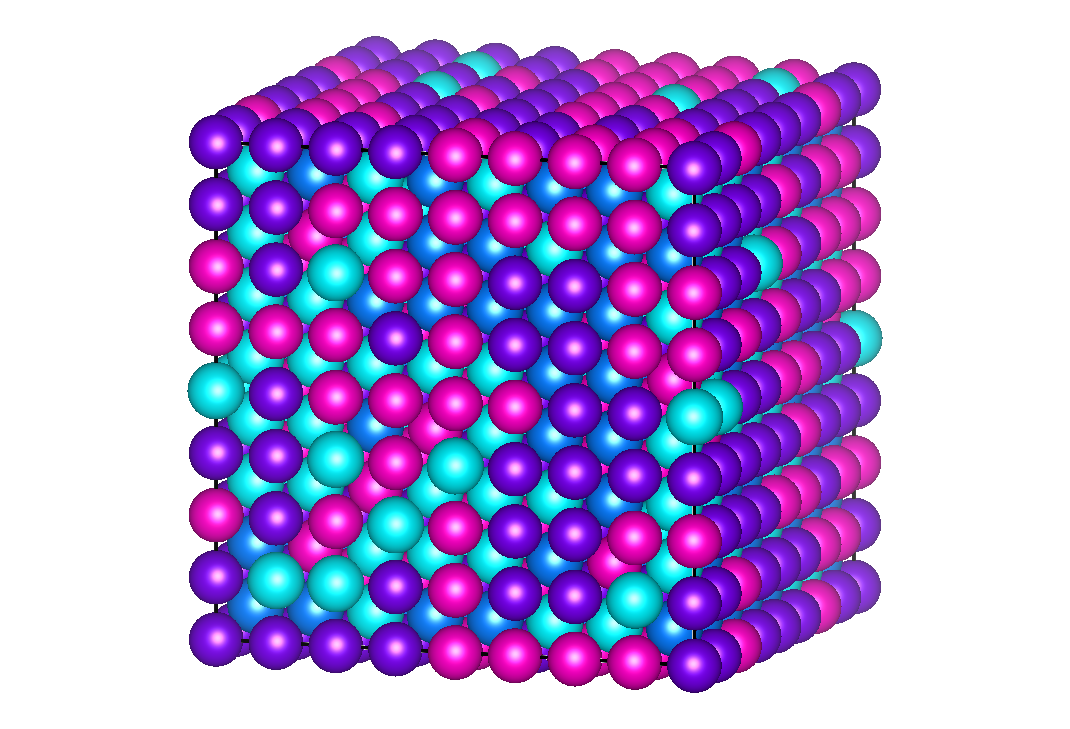}
  \includegraphics[width=.49\textwidth]{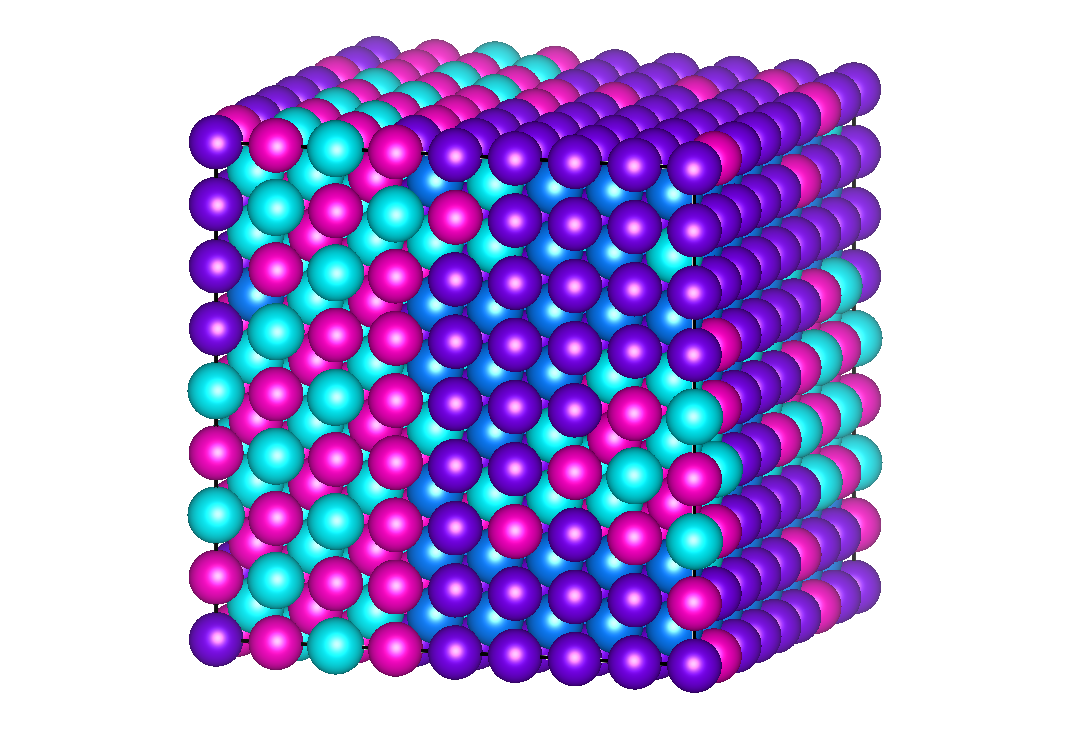}
  \caption{\label{fig:MoNbTaW-struct} $L=8$ quaternary MoNbTaW at $T=390$K (top) showing the B2 phase, and at $T=100$K (bottom) showing phase separation. The color scheme is chosen so that purple (Mo) and magenta (W) alternate with cyan (Nb) and blue (Ta). Lighter colors (NbW in B32 structure) segregate from dark colors (MoTa in B2 structure) at low $T$.}
\end{figure}

At temperatures below 300K the quaternary undergoes a second transition~\cite{Shapeev}, to a two-phase mixture of B2-type MoTa and B32-type NbW (see Fig.~\ref{fig:MoNbTaW-struct}). Full first principles calculations predict a more complex low temperature structure of Pearson type hR7~\cite{Widom2015}. Note that the separated structure is no longer single phase, so the CVM formulas do not apply. The specific heat and histogram entropy remain above zero, suggesting a possible incomplete equilibration at the lowest temperatures.

\section{Conclusions}

This work introduces an approach to modeling interatomic interactions based on a lookup table of precalculated cluster motifs. The interaction is quick to create from high throughput first principles total energy calculations. It can be readily generalized to a wide variety of high entropy alloys, and it scales well as the number of chemical species grows because the initial calculations are so fast and no subsequent fitting is required. We apply the model to replica exchange Monte Carlo simulations that efficiently sample the equilibrium ensembles across a broad range of temperatures. The data accumulated during the simulation is analyzed by multiple histogram methods that reveal thermodynamic properties as continuous functions of temperature. We also take simulated cluster frequencies as input to directly evaluate the entropy within the approximations of Kikuchi's Cluster Variation Method. Our computational tools are available in the public domain.

The results of our simulation confirm that the A2 to B2 transition lies in the Ising universality class, both for the binary MoTa and also for the quaternary MoNbTaW. Our results for the temperature-dependent entropy show that despite the presence of short-range order above $T_c$, as revealed by the pair frequencies $\yac$, the entropy loss is less than 20\%, around 0.1$\kB$ for MoTa and 0.2$\kB$ for MoNbTaW. Below $T_c$, with the onset of symmetry-breaking in the single-site occupation, the entropy drops rapidly. We explored the sequence of CVM-type approximations to the entropy and confirm that the best convergence is provided by the sequence site:NN:Tetrahedron. Lengthy simulations are required to obtain sufficient accuracy in the tetrahedron frequencies, so we suggest that stopping at the pair level should provide sufficient accuracy for most purposes.

\section{Acknowledgements}
ADK was supported by the SURF program at Carnegie Mellon University and by the National Science Foundation under grant number 2103958; MW was supported by the Department of Energy under Grant No. DE-SC0014506. We thank Wissam Al-Saidi and Fritz Koermann for useful discussions on interatomic interaction models.

\bibliography{refs}

\end{document}